%% file: main.tex
\documentclass{PoS}

\input{settings}
\title{Matching NLO QCD Corrections in WHIZARD with the POWHEG scheme}

\ShortTitle{Matching NLO QCD Corrections in WHIZARD with the POWHEG scheme}

\author{\speaker{Bijan Chokoufe Nejad}\\
        DESY Theory Group, Notkestr. 85, D-22607 Hamburg, Germany\\
        E-mail: \email{bijan.chokoufe@desy.de}}

\author{Wolfgang Kilian\\
        University of Siegen, Department of Physics,
        Walter-Flex-Str. 3, D-57068 Siegen, Germany\\
        E-mail: \email{kilian@hep.physik.uni-siegen.de}}

\author{J\"urgen Reuter\\
        DESY Theory Group, Notkestr. 85, D-22607 Hamburg, Germany\\
        E-mail: \email{juergen.reuter@desy.de}}

\author{Christian Weiss\\
        DESY Theory Group, Notkestr. 85, D-22607 Hamburg, Germany\\
        University of Siegen, Department of Physics,
        Walter-Flex-Str. 3, D-57068 Siegen, Germany\\
        E-mail: \email{christian.weiss@desy.de}}

\abstract{%
Building on the new automatic subtraction of NLO amplitudes in \whizard{}, we
present our implementation of the \powheg{} scheme to match radiative
corrections consistently with the parton shower.
We apply this general framework to two linear collider processes,
$e^+e^-\,\to\,t\bar{t}$ and $e^+e^-\,\to\,t\bar{t}H$.

\begin{flushright}
  \normalsize{} DESY 15--174
\end{flushright}
}

\FullConference{The European Physical Society Conference on High Energy Physics\\
                 22-29 July 2015\\
                 Vienna, Austria}

\newcommand\Sudakov\Delta

\newcommand\pt{p_{\T{T}}}

\begin{document}
\section{Introduction}
\label{sec:intro}
Polarized lepton colliders (LC) operated at high energies are indispensable
tools to increase the precision of various {SM} parameters and often complement
the measurements that are possible at hadron colliders like the LHC.
For numerous LC studies, the multi-purpose event generator
\whizard{}~\cite{0708.4233,1410.4505} is a commonly used simulation tool,
as it allows to study beamstrahlung as well as \ac{ISR} effects.
Moreover, high-multiplicity final states can be automatically and efficiently
generated using \OMega{}~\cite{hep-ph/0102195}.
So far, these predictions have been based on tree-level matrix elements combined
with conventional parton showers to describe the effects of QCD radiation.
To systematically improve this description, it is mandatory to include the
\ac{NLO} and avoid double counting with the parton shower.

The rigorous matching of \ac{NLO} computations with parton showers has been
pioneered with \mcatnlo{}~\cite{hep-ph/0204244}.
Its main principle is the subtraction of the expansion of the parton shower
from the cross section.
P. Nason proposed a similar method that avoids the inherent problem of
\mcatnlo{} of producing negative weight events,
in the sense that negative weights can only occur in regions where fixed-order
perturbation theory fails~\cite{hep-ph/0409146}.
Following the first implementation~\cite{hep-ph/0606275}, the algorithm has
been worked out in detail \cite{0709.2092} and dubbed the \powheg{} method
(Positive Weight Hardest Emission Generator).
As the hardest, \ie{} highest relative $p_T$, emission is not generated by the
attached parton shower but by the algorithm itself, it is guaranteed to maintain
the NLO accuracy of the sample, irrespective of the used shower.
This requires, though, that the shower respects the hardest emission, which is
easily satisfied with a \emph{veto} of higher $p_T$ on subsequent emissions.
In case the ordering variable of the shower is not $p_T$, soft radiation
before the hardest emission has to be added as well in terms of a
\emph{truncated} shower.

Following the work of \Rcite{0709.2092}, the semi-automated NLO+PS event
generator called \powhegbox{}~\cite{1002.2581} has been developed.
In this framework a multitude of LHC processes has been made publicly available.
The drawback of the \powhegbox{} is that it only automates parts of the
algorithm meaning that adding a new process requires considerable
theoretical effort from the construction of the phase space to the
implementation of the matrix elements.
With the advent of automated \acp{OLP} like \gosam{}~\cite{1404.7096} or
\openloops{}~\cite{1111.5206}, it has become feasible to build a fully
automated event generator using the \powheg{} method.
First steps in this direction were made in \sherpa{}~\cite{1008.5399}
as well as \herwig{}~\cite{1109.6256}.
In this work, we sketch the automation of the \powheg{} matching in
\whizard{}.
There also was earlier work on QED resummation matched to QED \ac{NLO}
calculations within \whizard{} leading to strictly positive-weight
events~\cite{hep-ph/0607127,0803.4161}.

\whizard{} has recently been augmented by an automation~\cite{WhizardNLO2015} of
the FKS subtraction~\cite{hep-ph/9512328,0908.4272}, which will be a key
ingredient in our discussion.
The impact of \powheg{} matching on event shapes at a
lepton collider like $e^+e^-\,\to\,\text{hadrons}$ has first been discussed
in \Rcite{hep-ph/0612281}, where a significant improvement in the description of the
measured data from LEP has been found in almost all observables, compared to
\ac{LO} with a matrix element correction.
In \Rcite{0812.3297}, this work has been extended to consider on-shell top-pair
production with semi-leptonic decays at the {ILC}.
Although \whizard{} spear-headed many beyond the SM phenomenological
studies~\cite{hep-ph/0411213,hep-ph/0512260,hep-ph/0604048,0806.4145,0809.3997,1408.6207},
we will focus here on SM QCD effects.
\section{\powheg{} matching}
\label{sec:powheg}
For completeness, we briefly sketch here how \powheg{} events are generated.
The corresponding proofs and more detailed information can be found in
\Rcite{hep-ph/0409146,0709.2092}.
Contrary to the subtractive approach of \mcatnlo{}, \powheg{} is a
unitary method to generate $n$- and $n+1$-parton event samples.
Hereby, we distribute the Born kinematics $\D\Phi_n$ according to the inclusive
NLO cross section
\begin{align}
  \bar{B} &= B + V + \Int{\D\Phi_\T{rad}} (R - C) \co{}
  \label{eq:bbar}
\end{align}
where $V=V_0+\Int{} C$ is the virtual part including the
analytically integrated subtraction terms $\int{C}$ that are subtracted again in
differential form from the real emission part $R$.
The integral over the radiation phase space $\D\Phi_\T{rad}$ in \cref{eq:bbar}
is evaluated numerically in a \ac{MC} sampling using \whizard{}'s standard
phase space integrator \prog{Vamp}~\cite{hep-ph/9806432} together with the
sampling over $\D\Phi_n$.
To obtain at least \ac{LL} correctness in $p_T$, we have to attach the
corresponding Sudakov form factors $\Sudakov\Al(\pt\Ar)$, yielding the
probability that no emission occurs between a high scale ${\pt}_\T{max}$ and
$\pt$
\begin{align}
  \Sudakov(p_T) &= \Exp{- \Int{\D\Phi_\T{rad}} \frac{R(\Phi_{\T{rad}})} {B}
        \Hs{k_T^2(\Phi_{\T{rad}}) - p_T^2}}\po
  \label{eq:sudakov}
\end{align}
With these quantities, we can write down the differential cross section as
\begin{align}
  \D\sigma &= \bar{B}\;\D\Phi_n \Cl(\Sudakov\Al(\pt^\T{min}\Ar) +
  \D\Phi_\T{rad}\Sudakov\Al(k_\T{T}(\Phi_{\T{rad}})\Ar)
  \frac{R(\Phi_{\T{rad}})}B\Cr) \po
  \label{eq:powhegxsec}
\end{align}
Note that the expression in parentheses in \cref{eq:powhegxsec} integrates to
one by the unitary construction as can be easily verified.
The first term corresponds to no emission down to ${\pt}_\T{min}$ and the second
to an emission at the scale $k_T$.
This ensures that the \ac{NLO} cross section is conserved, implying that the
\powheg{} matching only changes the spectrum.
Especially, it damps the emission of soft and collinear radiation of
the pure \ac{NLO} prediction since $\lim{\Sudakov(\pt\to0)}=0$.

The ratio $R/B$ in \cref{eq:sudakov} is the differential splitting probability 
and is approximated in parton showers by universal splitting kernels.
Using a process dependent ratio instead makes it significantly harder to
generate $\pt$ distributions according to this form factor.
There are two ways to circumvent this problem:
Obviously, one can use the universal properties of this ratio, \ie{} the known
soft and collinear divergence structure, to construct an overestimator $U$
weighted with a constant factor $N$.
Emissions are then accepted according to the probability $R/B/(NU)$.
A different approach is the fully numerical evaluation of the exponent in
\cref{eq:sudakov} as it is done in \exsample{}~\cite{1108.6182}.
In our implementation, we decided to use a hybrid version, where $N$ is a grid
that depends on the radiation variables multiplied with the general $U$
functions, similar to the approach in the \powhegbox{}.
A dedicated performance and validation comparison with \exsample{} featuring
multiple processes would be an interesting future project.

At this point, we want to stress the nice interplay between the FKS subtraction
and \powheg{} event generation.
While we have written \cref{eq:sudakov} in a general way, there are of course
different possible singular regions $\alpha$, each having a different emission
probability $R_\alpha/B$ with $R=\sum_\alpha R_\alpha$.
By having FKS at hand, we can directly retrieve these $R_\alpha$, which are
used to divide the real part into regions with at most one soft and one
collinear singularity.
Analogously, the overall Sudakov form factor will be a product of the
$\Sudakov_\alpha$ of the different regions.
In the implementation, this results to a generation of either one or no
emission in each region.
The region with the largest $\pt$ is kept in the event, also known as highest
bid method, \cf{} Appendix~B in \Rcite{0709.2092}.
\section{Effects of QCD radiation on top and electroweak physics at a future linear collider}
\label{sec:results}
%
In the following, jets are possible combinations of all occurring quarks and
gluons, clustered with \fastjet{}~\cite{1111.6097} according to an anti-$k_T$
algorithm that uses energies and spherical coordinates instead of transverse
momentum and rapidities as distance measure with $R=1.0$.
The shown events have been simulated only up to the first emission, leaving out
the subsequent simulation chain, in order to focus only on the \powheg{}
implementation.
It has been checked, though, that processing the \powheg{} events produced by
\whizard{} with \pythia{}'s $\pt{}$-ordered shower~\cite{1410.3012} in
the corresponding veto mode delivers reasonable physical results.

In our setup, the top mass is set to $m_t=\SI{172}{\GeV}$.
We chose $\mu_r=m_t$ as renormalization scale. The coupling constants are
$\alpha^{-1}=132.160$ with no running and $\alpha_S(M_Z)=0.118$ with a NLL
runnning with 5 active flavors.
\ac{LO} and \powheg{} events are unweighted during generation.
In the \ac{NLO} event samples, we associate Born kinematics with a weight of
$\mathcal{B} + \mathcal{V} - \sum_\alpha \mathcal C_\alpha$.
Together with this Born event, we generate for each singular region $\alpha$, a
real-emission event with weight $\mathcal{R}_\alpha$.
The histograms are generated with \rivet{}~\cite{1003.0694},
using \SI{500}{\K} {LO} and \powheg{} as well as \SI{1500}{\K} \ac{NLO} events.

\begin{figure}[htbp]
\centering
\includegraphics[width=0.45\textwidth]{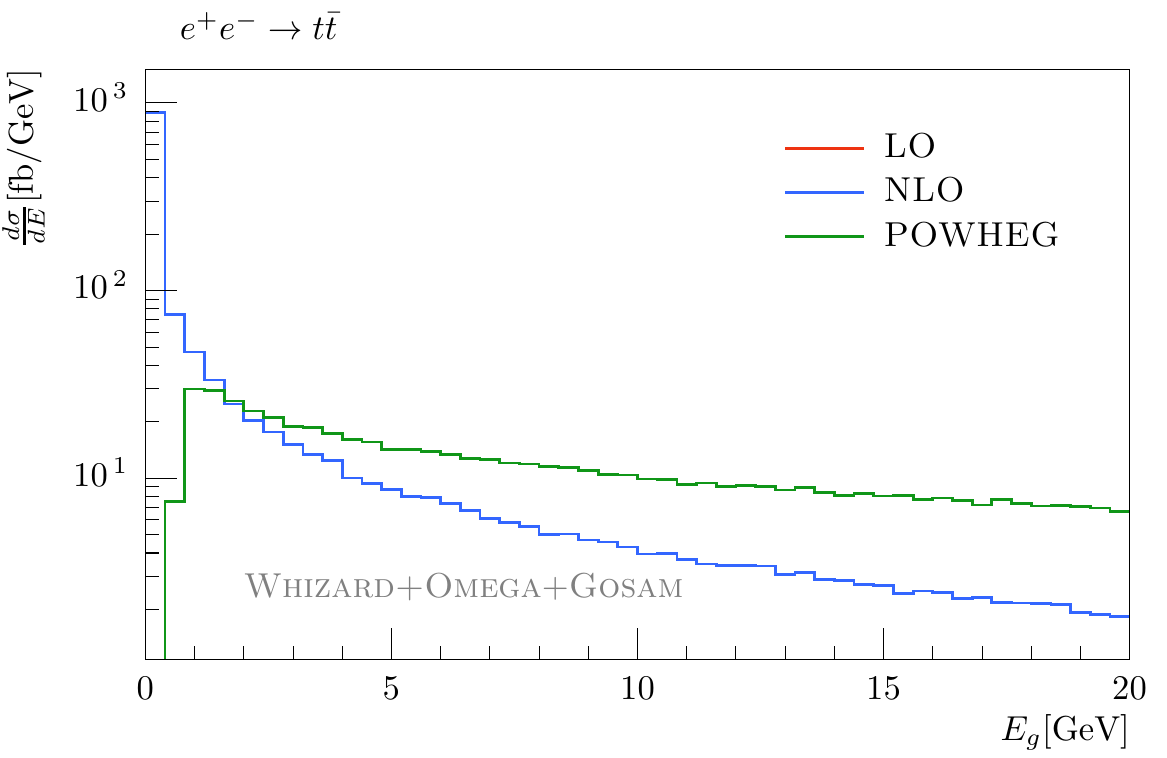}
\includegraphics[width=0.45\textwidth]{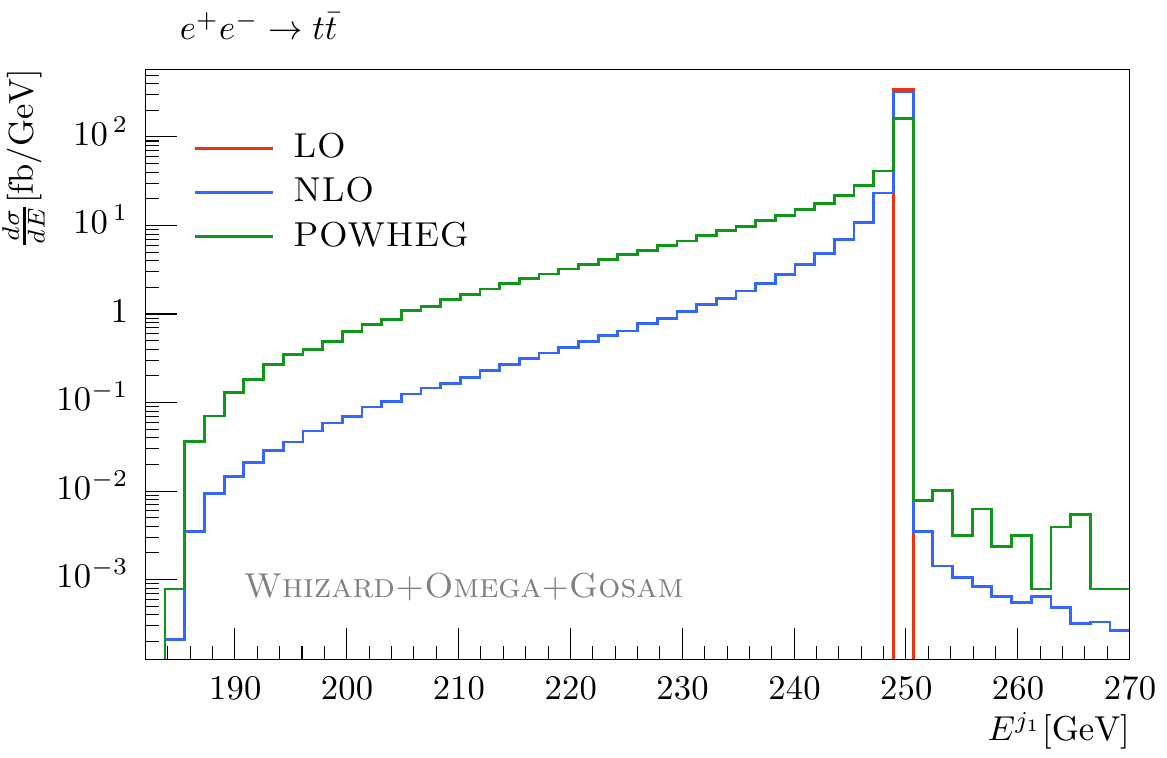}
\caption{Energy distributions of the emitted gluon and of the hardest jet.}
\label{fig:ttbar}
\end{figure}

\cref{fig:ttbar} shows onshell $t\bar{t}$ production at a lepton
collider with $\sqrt{s}=\SI{500}{\GeV}$.
Polarization and beamstrahlung effects as well as lepton ISR are neglected.
The soft gluon divergence can be seen in the \ac{NLO} event samples either
directly in the (unphysical) energy distribution of the gluon or indirectly in
the distribution of the hardest jet, which peaks around the Born value due to
mostly soft gluons.
By applying the Sudakov form factor, the \powheg{} events have the expected
suppression of this divergence.
Due to the unitary construction, this leads to an increase of the differential
cross section in the remaining part of the spectrum, a well known feature of
pure \powheg{} distributions.
As one might wish to restrict the effect of the Sudakov suppression to the area
where $\pt{}$ is small, the real radiation can be divided in a hard and a soft
part by means of \emph{damping factors}, bringing \powheg{} formally and
phenomenologically closer to \mcatnlo{}, while maintaining the benefits
discussed in \cref{sec:powheg}.
The associated freedom in the division between hard and soft part has to be
regarded as a theoretical uncertainty and will be discussed in a future work.

\begin{figure}[htbp]
\centering
\includegraphics[width=0.45\textwidth]{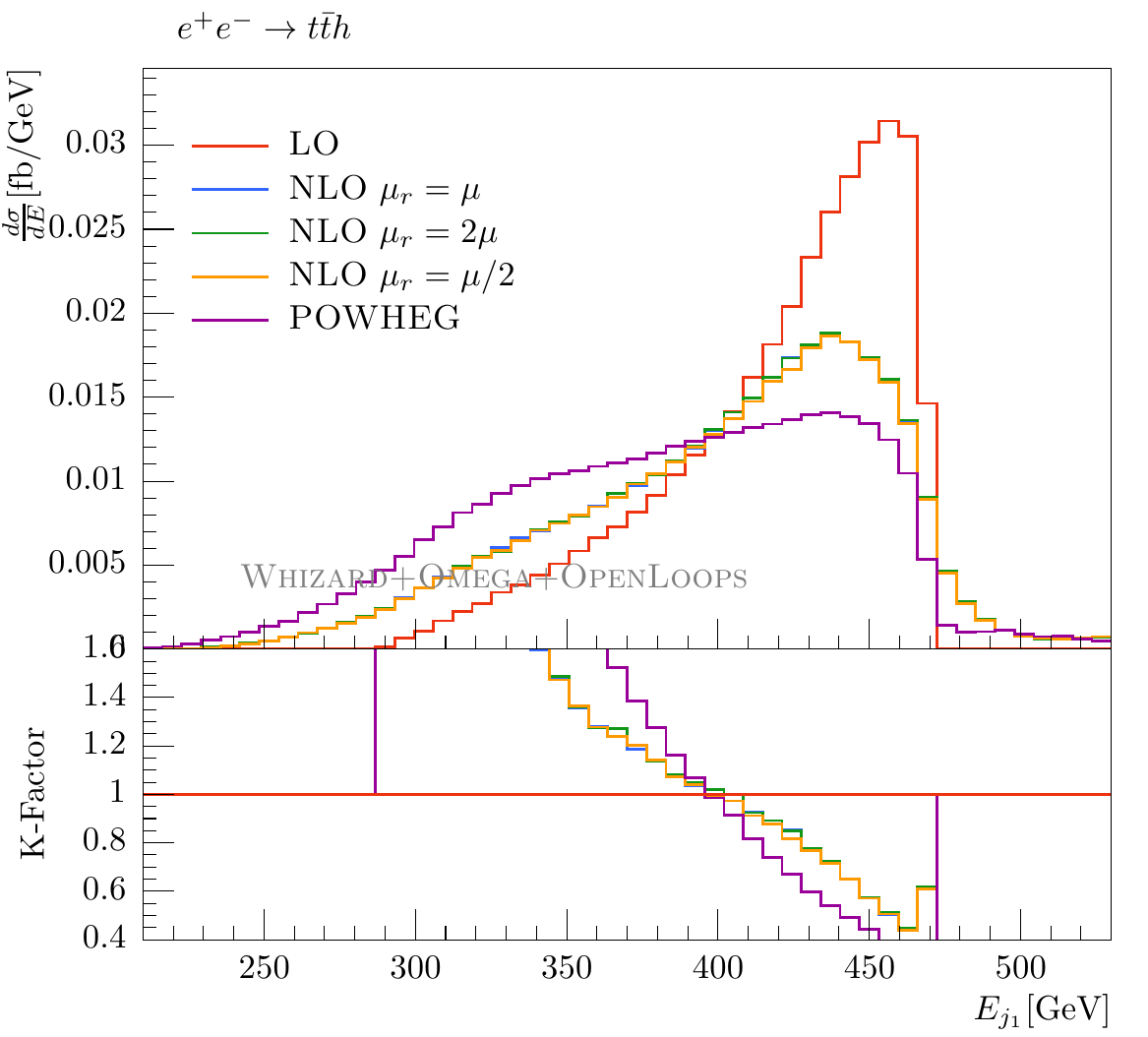}
\includegraphics[width=0.45\textwidth]{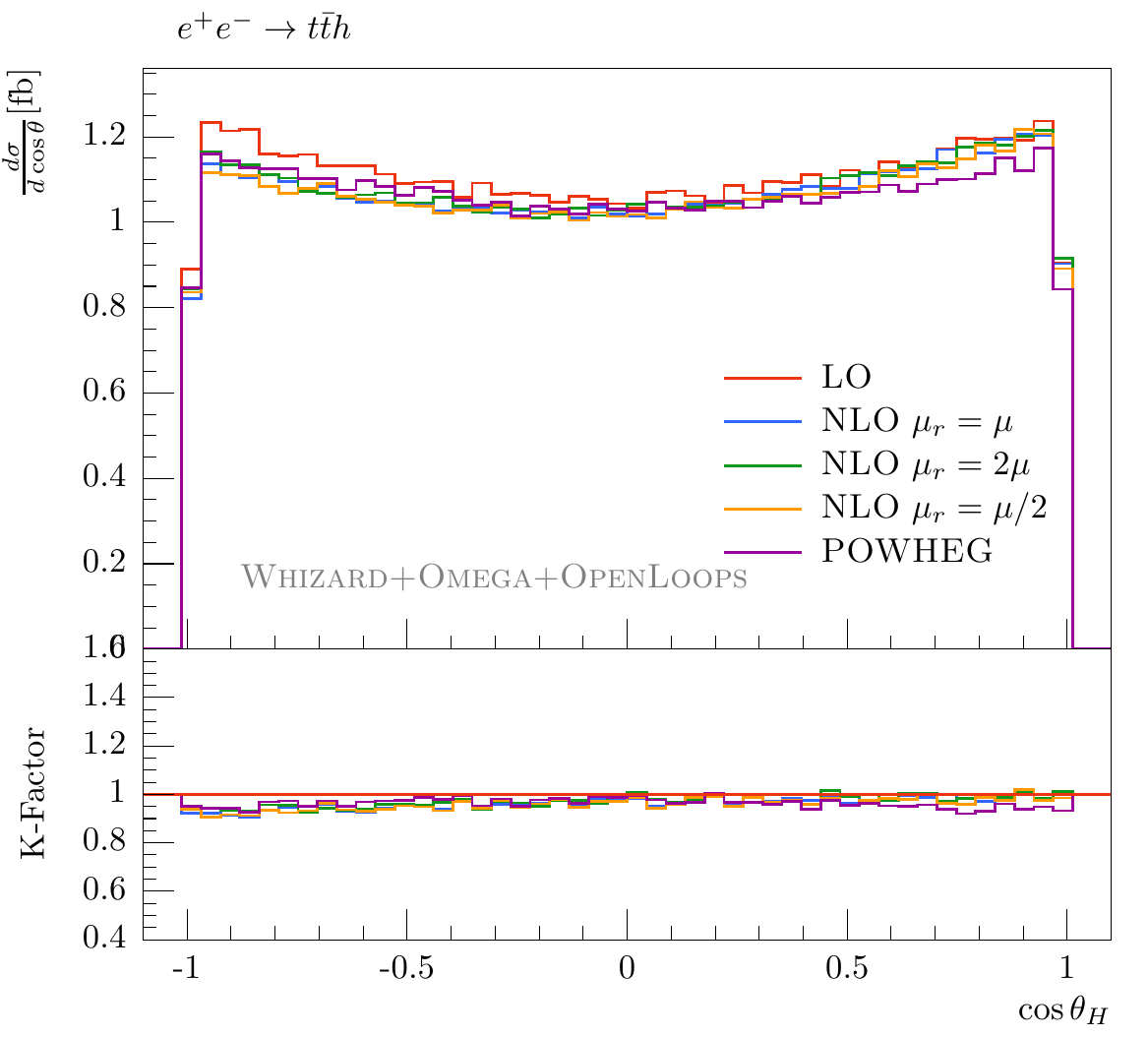}
\caption{The energy distribution of the hardest jet and the angular distribution
  of the Higgs boson.}
\label{fig:ttbarH}
\end{figure}

Let us now address $e^+e^-\,\to\,t\bar{t}H$ with the same setup as above but at
$\sqrt{s}=\SI{1000}{\GeV}$.
\cref{fig:ttbarH} shows distributions of two observables:
In the energy distribution of the hardest jet, we can see again the effect of
Sudakov suppression at the high energy peak.
For comparison, we also show the effect of scale variation, which, as expected,
does not cover the difference between \ac{NLO} and \powheg{}.
On the other hand, we observe that in inclusive quantities like the angular
distribution of the Higgs boson, the \powheg{} matching has no significant
effect.
This is of course only a cross check that inclusive quantities remain correct to
\ac{NLO}.
We find that although the total K-factor at this value of $\sqrt{s}$ is close to
1, distributions of observables that are sensitive to {QCD} radiation can change
drastically.
\section{Summary \& Outlook}
\label{sec:summary}
We have presented an independent implementation of the \powheg{} matching scheme
that builds on the recent automation of QCD NLO corrections in \whizard{}.
The key feature of the \powheg{} matching, namely the suppression of the
differential cross section for small relative $\pt$, has been reproduced and we
have shown for the first time the impact of the \powheg{} matching on
distributions for $e^+e^-\,\to\,t\bar{t}H$.
A more detailed analysis that focuses on the impact of damping factors and the
general assessment of the associated uncertainties in various processes will
follow.
Our implementation is process-independent but still subject to extensive
validation.
It can be tested for any lepton-collider process in the current publicly
available release of \whizard{}~\verb|2.2.7| but should be regarded as
experimental feature and results analyzed with care.
Hadron-collider processes are currently not supported but planned for the near
future.
\acknowledgments
We are grateful to N. Greiner for many helpful discussions on NLO event
generation in general as well as technical support with \gosam{}.
Also, we thank J. Lindert for providing a variety of before non-available
features in \openloops{}, especially the required loop libraries for the
processes studied in this paper.
We also appreciate the help of both of them for interfacing their respective
programs to \whizard{}.
Furthermore, we want to thank E. Bagnaschi for enlightening discussions on
\powheg{} and scales.
\printbibliography
\end{document}

%% file: settings.tex
\usepackage{amsmath}
\usepackage{acronym}
\usepackage{xspace}
\usepackage[hyphens]{url}
\usepackage{cleveref}
\usepackage{siunitx}
\usepackage{booktabs}
\usepackage{hepnames}

\DeclareSIUnit[number-unit-product = \;]{\permil}{\textperthousand}
\DeclareSIUnit[number-unit-product = \;]{\M}{M}
\DeclareBinaryPrefix\kibi{Ki}{10}
\DeclareBinaryPrefix\mibi{Mi}{20}
\DeclareBinaryPrefix\gibi{Gi}{20}

\newcommand\picpath{plots}
\DeclareMathAlphabet{\mathcal}{OMS}{cmsy}{m}{n}

\newcommand\Exp[1]  {\exp{\left\{#1\right\}}}

\newcommand\Int[2][]{\ems{\int#1#2\:}}

\newcommand\mr[1]{\mathrm{#1}}

\newcommand{\ems}[1]{\ensuremath{#1}}
\newcommand{\T}[1]{\mr{#1}}

\newcommand\po{\;.}
\newcommand\co{\;,}

\newcommand\Rcite[1]{Ref.~\cite{#1}}

\newcommand\ie{i.e.}

\newcommand\cf{cf.}

\newcommand\Al{\bigl}
\newcommand\Ar{\bigr}

\newcommand\Cl{\biggl}
\newcommand\Cr{\biggr}

\newcommand{\heaviside}[1]{\theta\left(#1\right)}

\crefformat{equation}{Eq.~(#2#1#3)}
\crefformat{section}{Section~#2#1#3}
\crefformat{figure}{Figure~#2#1#3}
\crefformat{table}{Table~#2#1#3}
\crefformat{chapter}{Chapter~#2#1#3}
\crefformat{appendix}{Appendix~#2#1#3}

\newcommand\prog[1]{\textsc{#1}}

\newcommand\fastjet{\prog{FastJet}\xspace}
\newcommand\rivet{\prog{Rivet}\xspace}
\newcommand\whizard{\prog{Whizard}\xspace}
\newcommand\OMega{\prog{O'Mega}\xspace}
\newcommand\pythia{\prog{Pythia8}\xspace}
\newcommand\sherpa{\prog{Sherpa}\xspace}
\newcommand\herwig{\prog{Herwig}\xspace}
\newcommand\powheg{\prog{Powheg}\xspace}
\newcommand\exsample{\prog{ExSample\xspace}}
\newcommand\mcatnlo{MC@NLO\xspace}
\newcommand\powhegbox{\prog{Powheg-Box}\xspace}
\newcommand\gosam{\prog{GoSam}\xspace}
\newcommand\openloops{\prog{OpenLoops}\xspace}

\newcommand{\D}{\ems{\T{d}}}

\newcommand{\Hs}[1]{\heaviside{#1}}

\newacro{1POW}{one-particle off-shell wave function}
\newacro{APS}{antenna pole structure}
\newacro{BSM}{beyond the SM}
\newacro{CAS}{Computer Algebra System}
\newacro{CERN}{European Organization for Nuclear Research}
\newacro{CMF}{center of mass frame}
\newacro{CPU}{central processing unit}
\newacro{CSE}{commmon subexpression elimination}
\newacro{FPGA}{Field programmable gate array}
\newacro{GPU}{graphics processing unit}
\newacro{HL-LHC}{High Luminosity LHC}
\newacro{HMC}{Helicity Monte Carlo}
\newacro{ILC}{International Linear Collider}
\newacro{ISR}{initial-state radiation}
\newacro{LCA}{leading color approximation}
\newacro{LL}{leading log}
\newacro{LHC}{Large Hadron Collider}
\newacro{LHS}{left-hand side}
\newacro{LO}{leading order}
\newacro{LSZ}{Lehmann-Symanzik-Zimmermann}
\newacro{MC}{Monte Carlo}
\newacro{MIC}{Many Integrated Cores}
\newacro{MSSM}{minimal supersymmetric SM}
\newacro{NLO}{next-to-leading order}
\newacro{NNLO}{next-to-next-to-leading order}
\newacro{NUMA}{Non-Uniform Memory Access}
\newacro{OLP}{One-Loop Provider}
\newacro{opcode}{operation code}
\newacro{OVM}{O'Mega virtual machine}
\newacro{PDG}{Particle Data Group}
\newacro{QCD}{quantum chromodynamics}
\newacro{QED}{quantum electrodynamics}
\newacro{RHS}{right-hand side}
\newacro{SIMD}{single instruction multiple data}
\newacro{SM}{standard model}
\newacro{VM}{virtual machine}

\usepackage[babel]{csquotes} 
\usepackage[style=numeric,
  maxnames=3,
  minnames=3,
  backend=bibtex,
  citestyle=numeric-comp,
  url=false,
  maxbibnames=5
  ]{biblatex}
\DeclareFieldFormat{journaltitle}{\emph{#1},}
\DeclareFieldFormat[inbook,thesis]{title}{{#1}\addperiod}
\DeclareFieldFormat[article]{title}{\mkbibemph{#1}}
\DefineBibliographyStrings{english}{%
    backrefpage  = {see p.}, 
    backrefpages = {see pp.} 
}
\DeclareFieldFormat{eprint}{%
  \iffieldundef{eprintclass}
    {}
    {\thefield{eprintclass}/}%
    #1%
}
\DeclareFieldFormat{eprint:arxiv}{[\printfield{eprint}]}
\DeclareFieldFormat[article]{volume}{{#1}\addcolon\space}
\AtEveryBibitem{\clearfield{month}}
\AtEveryBibitem{\clearfield{title}}

\renewbibmacro{in:}{}